Anisotropy of shear relaxation in primitive chain network simulations for entangled polymers confined between flat walls

Yuichi Masubuchi

Department of Materials Physics, Nagoya University, Nagoya 4648603, JAPAN

**ABSTRACT**

Anisotropic shear relaxation is an interesting but rarely discussed issue in polymer dynamics under confinement [Abberton et al., Macromolecules, 48, 7631, 2015]. According to the earlier study of bead spring simulations for an unentangled polymer melt confined between two flat plates, the shear relaxation modulus taken perpendicular to the interface is accelerated by decreasing the distance between plates, whereas the parallel component is unchanged. This study observed similar anisotropic shear relaxation for entangled polymer melts under confinement in multi-chain slip-link simulations (primitive chain network simulations). The analysis demonstrated that the accelerated relaxation in the perpendicular component reflects the Rouse-type constraint release dynamics, for which the coarsening is upper-limited by the geometry. This result suggests a novel mechanism for anisotropic shear relaxation different from the modified chain statistics under confinement considered for unentangled systems.

**INTRODUCTION**

Due to its industrial and scientific significance, many attempts have been made to the effect of confinement on polymer dynamics. The polymer motion is affected by interfacial interactions and geometrical constraints. For thin polymer films with a free surface not attached to a solid substrate, Tg decreases with decreasing the film thickness, reflecting enhanced mobility partly due to the condensation of the chain ends[1–6]. In contrast, near the solid interface with attractive interaction, segment motion is suppressed[7,8]. Besides, geometrical confinement modifies the chain statistics [9–12]. The interfacial effects propagate in a long distance to the bulk,[13] and their contributions strongly depend on experimental conditions such as annealing of the specimen[14]. As a result of correlated contributions among these different mechanisms, diverse and seemingly contradictory results for the effect of confinement on the polymer dynamics have been reported[15–21]. To consider various factors separately and observe molecular conformation directly, molecular simulations have also been performed[22–26]. For instance, modification of the entanglement network near the surface has been discussed [25,26].

A not frequently discussed issue in confined polymer dynamics is anisotropic shear relaxation. Abberton et al.[27] performed bead-spring molecular dynamics simulations for polymers confined between smooth surfaces to report that shear relaxation modulus $G(t)$ exhibits different behavior depending on the direction according to the surface. The modulus measured under shear parallel to the surface, the so-called in-plane relaxation $G_\parallel(t)$, does not depend on the width between the solid walls, $h$, for the conditions $1.6 \lesssim h/R_g \lesssim 77$, whereas the modulus perpendicular to the surface, the so-called out-of-plane relaxation $G_\perp(t)$, strongly depends on $h$, being accelerated with decreasing $h$. Because their chains are unentangled (with a bead number per chain of 85), they analyzed their results with the Rouse model, incorporating the anisotropic chain statistics semi-empirically. Kirk and Ilg [28] proposed another modified Rouse model, in which the bond potential between Rouse beads is changed according to the Kuhn segment statistics near the surface due to the reflective boundary condition. They extended this idea to entangled systems by using the single-chain slip-spring model[29].

Nevertheless, anisotropic shear relaxation has never been discussed for entangled systems. Even for the extended single-chan slip-spring model by Kirk et al.[29], no anisotropic relaxation is expected because the modification in the Rouse spring is introduced as a function of distance from the surface and is not dependent on orientation. Their arguments imply that shear relaxation for entangled systems is probably isotropic even under confinement. However, in the earlier models, the position of entanglement is fixed in space and immobile, and the situation may change if thermal fluctuations of entanglements are considered. Such fluctuations overwhelm other relaxation mechanisms in a length scale smaller than the polymer dimension, and they become dominant for confined systems.

In this study, primitive chain network (PCN) simulations were performed for entangled melts confined between flat surfaces, and $G_\parallel(t)$ and $G_\perp(t)$ were measured. The results show that, similarly to the unentangled case reported earlier, $G_\parallel(t)$ is insensitive to $h$, whereas $G_\perp(t)$ is accelerated with decreasing $h$. The longest relaxation time of $G_\perp(t)$ demonstrates that the acceleration reflects the Rouse relaxation of blobs with their size of $h$. Details are shown below.

**MODEL AND SIMULATIONS**

The model examined in this study is a multi-chain slip-link model (the PCN model) for which the dynamics of several entangled polymer systems have proven consistent with experimental results[30–75]. Entangled polymer chains are replaced by a network consisting of nodes, strands, and dangling ends, and each polymer chain corresponds to a path connecting two dangling ends through several strands. Each network strand carries some Kuhn segments comparable to the entanglement segment defined in the tube segment. The chains are bundled in pairs at the network nodes by slip links, which are created and annihilated at the chain ends due to the reptation motion of the chains. The position of slip links (i.e., network nodes) and dangling ends obeys the Langevin equation of motion, in which the force balance is considered among drag force, tension acting on diverging strands, osmotic force, and thermal random force. The number of Kuhn segments on the strands develops according to the change rate equation where the same force balance as the slip link motion is considered along the chain. This transport of the Kuhn segments corresponds to the reptation dynamics of the chain through the slip links, and it induces disentanglement and hooking of another chain at the chain ends. Namely, when the number of Kuhn segments at a dangling end exceeds a critical value, the end segment hooks another segment randomly chosen from the surroundings to create a new network node with a slip link. Vice versa, when the Kuhn segment number becomes lower than the other critical value, the connecting node and the slip link are removed to release the hooked chain. Units of length, energy, and time were the average strand length $a$, thermal energy $kT$, and diffusion time of single network node $\tau = \zeta a^2/6kT$, where $\zeta$ is the friction constant.

The confinement was introduced via reflective boundary conditions for two parallel planes of the simulation box in a similar manner proposed earlier by Okuda et al.[76] For the other planes, periodic boundary conditions were applied. Figure 1 shows a typical snapshot of a polymer melt containing monodisperse linear polymers with the number of entanglement segments per chain $Z = 40$ in the unconfined bulk state. The boundary conditions are reflective in the compressed direction and periodic in the other directions. Note that, unlike the earlier study[70], no surface-grafted chains exist.

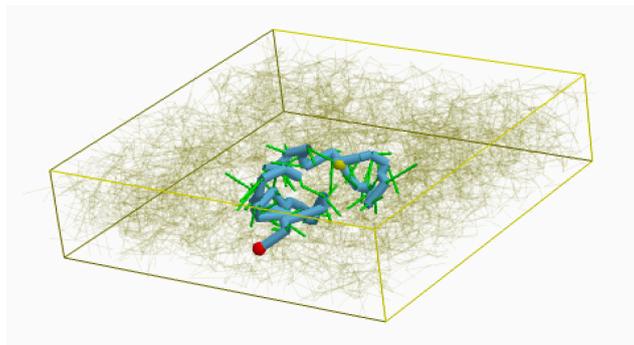

**Figure 1** (Color online) Typical snapshot of the primitive chain network model confined between flat surfaces at both compression sides. The molecular weight of each chain is $Z = 40$, and one of the included chains is highlighted. Red and yellow balls indicate both chain ends. The simulation box size is 4 and 16 in the compressed and the other directions, respectively.

The simulations were conducted for polymer melts with $Z = 10$, 20, and 40. The simulation box dimension $h$ was varied from 4 to 32 in the direction normal to the reflective flat surfaces (i.e., the compression direction in Fig 1) and chosen at 16 for the other directions. For statistics, eight independent simulation runs were performed for each condition. The relaxation functions were obtained according to the Green-Kubo relation from the trace of simulations under equilibrium taken for a sufficient period, which is at least 10 times longer than the longest relaxation time of the system. $G_\parallel(t)$ and $G_\perp(t)$ were obtained from the shear components parallel and normal to the flat surface of the stress tensor. The end-to-end relaxation functions $P_\parallel(t)$ and $P_\perp(t)$ were also acquired.

Note that Okuda et al.[76] examined similar systems and reported that i) the chain dimension decreases with decreasing $h$ in the compressed direction, whereas it is unchanged in the directions parallel to the surface, and ii) the chain diffusion and the end-to-end relaxation (on average for all the directions) are slightly accelerated by the compression, reflecting a weak reduction of entanglement density near the surface. These results are rationalized by the fact that the PCN model does not explicitly consider excluded volume effects. Because these behaviors were commonly observed for this study, chain dimensions and diffusion are not reported below.

**RESULTS**

Figure 2 shows the obtained $G_\parallel(t)$ and $G_\perp(t)$ for $Z = 20$ with various $h$. Here, the relaxation functions are normalized by their values at $t = 0$, described as $G_{\parallel 0}$ and $G_{\perp 0}$. Overall, the $h$-dependence of these relaxation functions is qualitatively similar to that reported for the unentangled system[27]. Namely, $G_\parallel(t)$ (broken curves) is virtually insensitive to $h$ and close to $G(t)$ obtained for the bulk liquids. In contrast, $G_\perp(t)$ (solid curves) is accelerated with decreasing $h$.

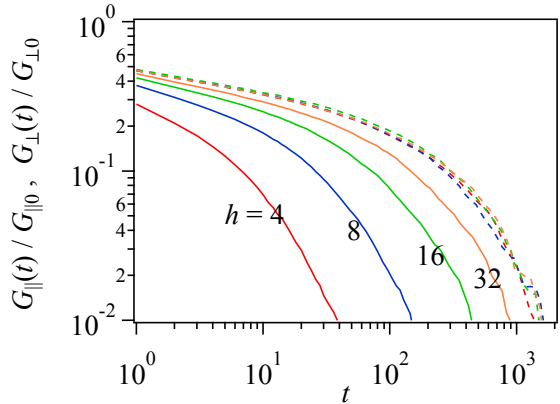

**Figure 2** (Color online) $G_\parallel(t)$ (broken curves) and $G_\perp(t)$ (solid curves) for $Z = 20$ with various

$h$ ranging from 4 to 32. The relaxation moduli are normalized by their values at $t = 0$.

Figure 3 shows the end-to-end relaxation functions $P_\parallel(t)$ and $P_\perp(t)$ for $Z = 20$ with various $h$. The $h$-induced acceleration is also observed but not significant compared to the modulus. One may argue that this result contradicts the earlier study by Okuda et al.[76], who reported insensitivity of the longest end-to-end relaxation time to confinement. However, these results are consistent because the slowest relaxation mode for the chain as a whole reflects $P_\parallel(t)$. Indeed, the end-to-end relaxation for the entire chain $P(t)$ almost overlaps with $P_\parallel(t)$ (not shown).

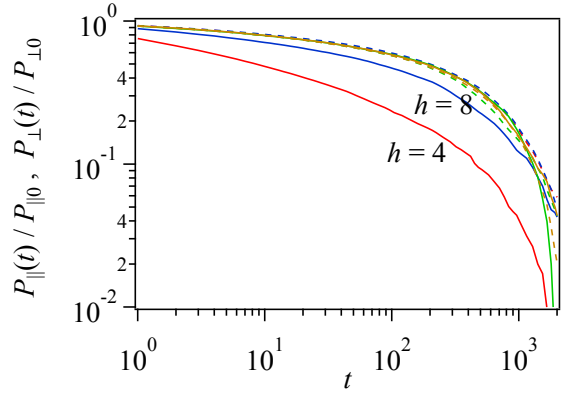

**Figure 3** (Color online) $P_\parallel(t)$ (broken curves) and $P_\perp(t)$ (solid curves) for $Z = 20$ with various $h$ ranging from 4 to 32. The relaxation functions are normalized by their values at $t = 0$.

Figure 4 indicates the longest relaxation time extracted from $G_\perp(t)$ (panel a) and $P_\perp(t)$ (panel b) plotted against $h$. As observed for the unentangled case, the relaxation time for modulus, $\tau_{G\perp}$, increases with increasing $h$. Note that Abberton et al.[27] fitted the relaxation function by the stretched exponential function to conclude that the relaxation time is not sensitive to $h$. Here, $\tau_d$ was determined via procedure X[77], and a multi-mode Maxwell relaxation function is assumed for the fitting. At a large $h$, as expected, $\tau_{G\perp}$ saturates to the value observed for the bulk liquid. However, the saturation occurs at $h$ significantly larger than $R_g$ and comparable to the contour length. Another interesting feature for $\tau_{G\perp}$ is that, before the saturation, $\tau_{G\perp}$ is independent of $Z$ and sorely depends on $h$. As the broken line indicates, $\tau_{G\perp} \propto h^2$. In contrast, the relaxation time for the end-to-end vector, $\tau_{P\perp}$, does not significantly depend on $h$, and only slightly decreases with decreasing $h$ when $Z$ is large.

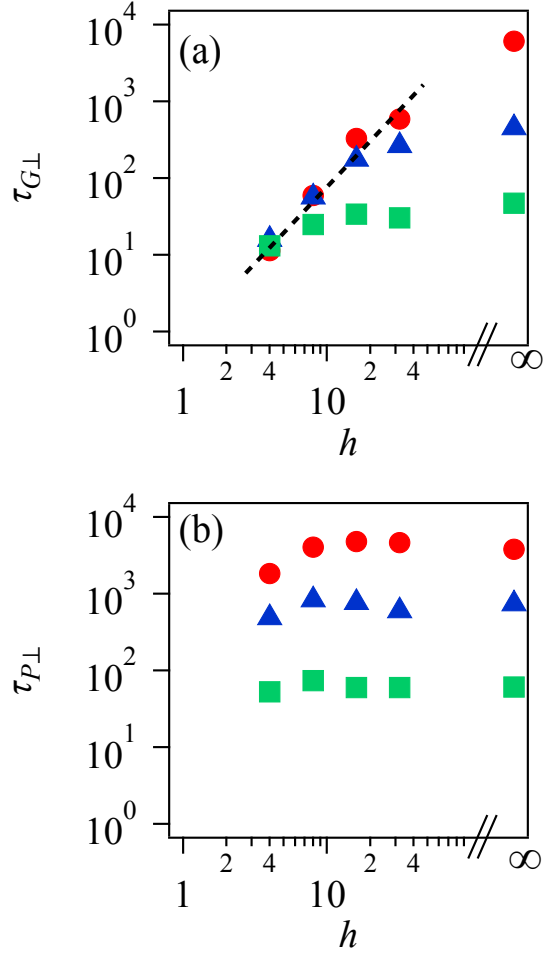

**Figure 4** (Color online) The longest relaxation time for $G_\perp(t)$ (a) and $P_\perp(t)$ (b) plotted against $h$ for $Z$=10 (green square), 20 (blue triangle), and 40 (red circle). The rightmost symbols indicate the values for the bulk liquids. The broken line in panel (a) shows a slope of 2.

**DISCUSSION**

A question arising from the abovementioned results is why $G_\perp(t)$ is accelerated. A possible explanation is that $G_\perp(t)$ shown in Fig 2 is the geometrical average from different layers with different distances from the surface. This argument is tested in Figure 5, in which $G_\perp(t)$ for the surface layer and bulk layers are separately shown. Here, the surface layer is defined in the area where the distance from the surfaces is smaller than 2. $G_\perp(t)$ for bulk layers $G_{b\perp}(t)$ (broken curves in the upper panel) is not different from that taken for the entire system (solid curves). Since the surface layer is considered for upper and lower surfaces, for the case with $h = 4$ no bulk layer exists. In contrast, $G_\perp(t)$ for surface layers $G_{s\perp}(t)$ (dashed curves in the lower panel) does not depend on $h$, except for $h = 4$. Nevertheless, it so appears that $G_\perp(t)$ shown in Fig 2 is the response from the bulk liquids.

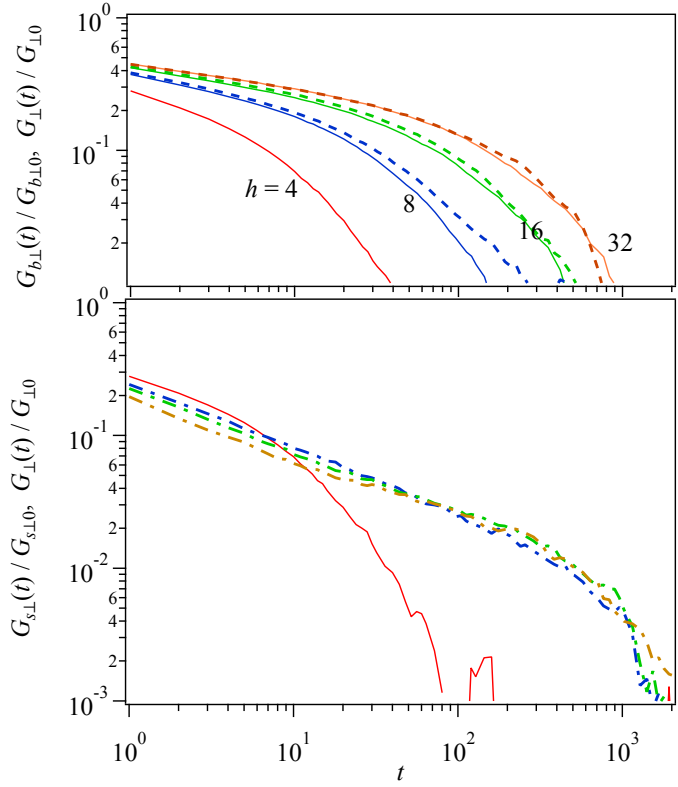

**Figure 5** (Color online) $G_\perp(t)$ separately obtained for the layers in the bulk $G_{\perp b}(t)$ (upper panel) and near the surface $G_{\perp s}(t)$ (lower panel) for $Z = 20$. The results for $h = 4, 8, 16,$ and $32$ are shown by red, blue, green, and orange broken and dashed curves, respectively. Solid curves are $G_\perp(t)$ for the entire system.

Another explanation is that the geometry enhances fast modes and smears slower modes. Let us consider a blob composed of several entangled subchains, and its size is comparable to $h$. The fundamental assumption underlying this picture is that local equilibration occurs within each blob. The PCN model is consistent with this setting due to the force balance around network nodes[78,79]. Let us further assume that $G_\perp(t)$ reflects such local relaxation within the blob. To examine this blob argument, Figure 5 compares $G_\perp(t)$ obtained for $Z=20$ and 40 at $h = 4, 8,$ and 16, demonstrating that $G_\perp(t)$ is independent of $Z$ and determined by $h$ when the chain is sufficiently larger than confinement. From Fig 4 (a), the blob relaxation time (drawn by the broken line) is written as $\tau_b \approx h^2$. Assuming that the blob molecular weight is $Z_b^2 = h^2$, one can have a relation $\tau_b \approx Z_b^2$. This relation implies Rouse-type dynamics. However, for the PCN model, the Rouse relaxation time of a chain with a given $Z$ is written as $\tau_R = Z^2/2\pi^2$, which is much smaller than $\tau_b$. Indeed, $\tau_b/\tau_R(Z_b) = 2\pi^2$. This difference implies that the relaxation inside the blob is not the Rouse relaxation of unentangled subchains but is attributable to the Rouse-type constraint release[80,81], for which the geometry limits the fat tube diameter. Note that the blob discussed here differs from that defined by Brochard and de Gennes[82] since in the PCN model, hydrodynamic force is not considered, and osmotic force is rather weak.

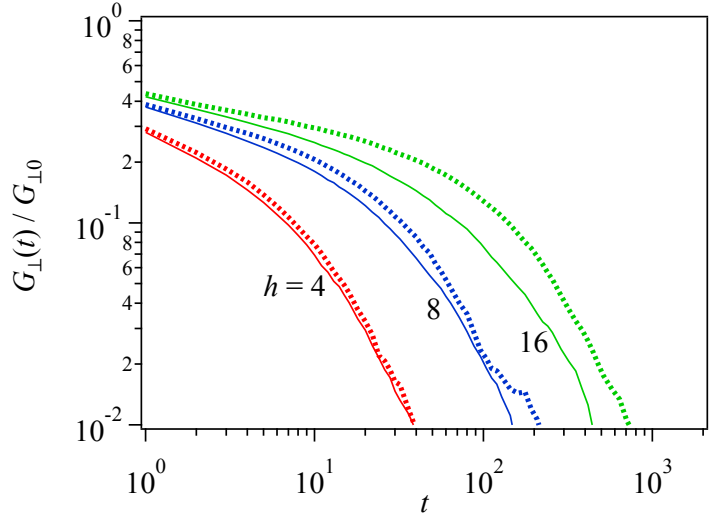

**Figure 6** (Color online) $G_\perp(t)$ at $h = 4$ (red), 8 (blue), and 16 (green) for $Z = 20$ (solid) and 40 (dotted).

## CONCLUSIONS

Primitive chain network simulations were performed for entangled polymer melts confined between flat walls without specific interfacial interactions. The shear relaxation modulus was obtained for directions parallel and perpendicular to the surface. Being similar to the earlier work for an unentangled system, the perpendicular component $G_\perp(t)$ is accelerated with decreasing the distance between walls $h$, whereas the parallel component $G_\parallel(t)$ is insensitive to $h$. Additional simulations with various molecular weights $Z$ demonstrated that $G_\perp(t)$ is independent of $Z$ and dependent only on $h$ unless the chain dimension is significantly smaller than $h$. The relaxation time of $G_\perp(t)$, $\tau_{G\perp}$, is proportional to $h^2$, implying that $G_\perp(t)$ reflects Rouse-type constraint release dynamics.

It should be emphasized that the examined model does not consider the modified chain statistics in the interface. For such a problem, the effects of interfacial and excluded volume interactions should be discussed via bead-spring simulations. Such supplemental studies are ongoing, and the results will be published elsewhere.

## ACKNOWLEDGEMENTS

The author appreciates the kind invitation from Prof. Keiji Tanaka and Prof. Satoru Yamamoto at Kyushu University to submit this study to the journal. The author also thanks for the helpful discussion with Prof. Tanaka. This study was partly supported by JST-CREST (JPMJCR1992) and JSPS KAKENHI Grant Number 22H01189.